\documentclass[twocolumn]{aastex62}

\usepackage{textcomp}
\usepackage{amsmath}
\usepackage{float} 

\usepackage{lineno}

\received{14.May 2020}
\accepted{24.Sept 2020}

\submitjournal{Icarus}

\begin{document}

\title{Cosmic radiation does not prevent collisional charging in (Pre)-Planetary Atmospheres}

\correspondingauthor{Felix Jungmann}
\email{felix.jungmann@uni-due.de}

\author{Felix Jungmann}
\affil{University of Duisburg-Essen, Faculty of Physics, Lotharstr. 1-21, 47057 Duisburg, Germany}

\author{Tetyana Bila}
\affil{University of Duisburg-Essen, Faculty of Physics, Lotharstr. 1-21, 47057 Duisburg, Germany}

\author{Laura Kleinert}
\affil{University of Duisburg-Essen, Faculty of Physics, Lotharstr. 1-21, 47057 Duisburg, Germany}

\author{Andre M\"olleken}
\affil{University of Duisburg-Essen, Faculty of Physics, Lotharstr. 1-21, 47057 Duisburg, Germany}

\author{{Rolf M\"oller}}
\affil{University of Duisburg-Essen, Faculty of Physics, Lotharstr. 1-21, 47057 Duisburg, Germany}

\author{Lars Schmidt}
\affil{University of Duisburg-Essen, Faculty of Physics, Lotharstr. 1-21, 47057 Duisburg, Germany}

\author{Niclas Schneider}
\affil{University of Duisburg-Essen, Faculty of Physics, Lotharstr. 1-21, 47057 Duisburg, Germany}

\author{Jens Teiser}
\affil{University of Duisburg-Essen, Faculty of Physics, Lotharstr. 1-21, 47057 Duisburg, Germany}

\author{{Detlef Utzat}}
\affil{University of Duisburg-Essen, Faculty of Physics, Lotharstr. 1-21, 47057 Duisburg, Germany}

\author{Victoria Volkenborn}
\affil{University of Duisburg-Essen, Faculty of Physics, Lotharstr. 1-21, 47057 Duisburg, Germany}

\author{Gerhard Wurm}
\affil{University of Duisburg-Essen, Faculty of Physics, Lotharstr. 1-21, 47057 Duisburg, Germany}


\begin{abstract}
In (pre)-planetary environments, dust and sand grains regularly collide. They electrically charge and discharge during these events. In this work, we study if cosmic radiation has any influence on the equilibrium charge state on timescales of minutes. We developed an experiment that was carried out during the ascent of a stratospheric balloon. With increasing altitude, the radiation activity increases by a factor of 54. However, we found only a very minor decrease in grain charges of up to 30\%. This implies that charge-moderated processes from thunderstorms on Earth, over early phases of planet formation to particle motion on the Martian surface on short timescales essentially proceed unhindered from a direct influence of cosmic radiation.

\end{abstract}


\section{Introduction}

The project reported here was originally motivated by two recent works by \citet{Steinpilz2020a} and \citet{Steinpilz2019}. In the first paper, \citet{Steinpilz2020a} found in drop tower experiments that collisional charging can promote aggregation in early phases of planet formation. Similar to the results of the ground-based experiments of \cite{lee2015} differently charged particles are able to form stable aggregates. Since collisional charging seems to be omnipresent in various occasions \citep{poppe2000,cimarelli2014} it is quite logical that charge-moderated aggregation can have an influence on planet formation. In the second paper, \citet{Steinpilz2019} give first data from a space station experiment, indicating grains to be less charged. 
\citet{Steinpilz2019} speculated that the higher cosmic radiation levels in the space station environment might be responsible for less charge on the grains. Since not only the net charge of the grain's surface might play an important role during aggregation but also the dipoles and multipoles \citep{matias2018}, local changes of charge can have a crucial impact on aggregation.  If cosmic radiation  is able to neutralize charge, it might also have consequences for quite different fields apart from planet formation. 

The most obvious and immediate application would be to collisional charging within our planet’s atmosphere. If charge levels varied between the ground and the top of the troposphere, in consequence, it would influence the formation of thunderstorms as they might at least in parts rely on collisional charging of ice grains \citep{Saunders2008}. It might also be important at the surface of Mars. This planet only owns a thin atmosphere and has an insignificant magnetic field on planetary scale, insufficient to protect it from energetic charged particles \citep{Zeitlin2019}. This results in high levels of cosmic radiation on the ground. Nevertheless, also here, collisional charging is discussed, e.g. in the context of particle transport at the surface \citep{Harrison2016}.

We refrain from trying to place our work in detailed context in all these applications. This would make this introduction very heterogeneous as these fields are usually well separated and this would go far beyond this work. In view of the results presented below, this would also be excessive. Instead, we will just concentrate on the basic question studied, namely, if collisional charging is immediately influenced by the different cosmic radiation levels present in these environments.

But why should cosmic radiation have an impact on the charge state of colliding grains in the first place? 
There are several aspects, that come to mind and that might not be discarded \textit{a priori}. 
First, cosmic radiation provides free charges one way or the other. These might simply discharge grains or charge them on a lower level. In planetary applications on airless bodies, grains are often considered to be charged and discharged by radiation \citep{Colwell2007}.
 
Second, several charging mechanism are discussed in the literature from trapped electrons in excited energy levels to exchange of ions \citep{Lacks2011}. Radiation can influence all of them, i.e. by populating the higher electron states or ionizing the water at a grain's surface. Apart from this, cosmic radiation can traverse grains and change the internal charge distribution, not only the surface. And last, in a gaseous environment, charging is limited by breakthrough voltages between grains \citep{Wurm2019, Harper2016}. Also here, ionizing radiation or ions provided by the radiation might trigger discharges more easily.

As first approach to study if cosmic radiation really has any influence, we built an experiment where grains are charged in collisions in a varying radiation environment.

\section{Radiation environments}

The idea behind this work is very simple - measuring the charge on colliding grains depending on different levels of radiation. There are quite different radiation environments though. Any effect might depend on the gas (species and pressure), on the kind of radiation ($\gamma, \beta, \alpha$, protons, ...), on the radiation energy (eV to GeV), or on the radiative flux \citep{Herbst2019a}.

Systematic studies in ground based laboratories with specific radiation parameters might be useful. However, as a first task we were interested in the question if radiation of the kind relevant for protoplanetary disks or planetary atmospheres would change collisional charging. Cosmic radiation is not purely of one kind and energy but is a mix generated by different processes from primary $\gamma$ particles or protons at different energies to secondary cascades of deeply penetrating muons and electrons. 
Using natural (cosmic) radiation might be the simplest but also a very appropriate approach to simulate (pre)-planetary environments.

The similarity of cosmic radiation on Earth and in protoplanetary disks might not be stressed too much but there are some common concepts.
The surface densities of the atmosphere on sea level on Earth and in the midplane of protoplanetary disks at 1 AU are on the same order of magnitude. In the minimum mass solar nebula it is 17000 $\rm kg/m^2$ \citep{Hayashi1985}. On Earth, it is 9800 $\rm kg/m^2$ (calculated from Earth's fact sheet at nssdc.gsfc.nasa.gov). What prevents charged cosmic radiation to enter Earth’s atmosphere too deeply are magnetic fields of about 50 $\rm \mu T$. Such magnetic field strengths are also present in protoplanetary disks \citep{Bertrang2017,Dudorov2014,Brauer2017}. 

The ionization rate in the dense midplane of protoplanetary disks might be as low as $\alpha = 10^{-23} s^{-1}$ and not larger than  $10^{-18} s^{-1}$ throughout the disk \citep{Cleeves2013}. On Earth, the number of ions produced by cosmic radiation varies from about  $\rm I = 1\, cm^{-3} s^{-1}$ on ground level to a maximum of about 1000 $\rm cm^{-3} s^{-1}$ at the Pfotzer maximum (16-20 km altitude) in special high energy events \citep{Herbst2019a}. Therefore, on ground we have an ionization rate of $\alpha =  5 \cdot 10^{-20} s^{-1}$ increasing with height to about $5 \cdot 10^{-15} s^{-1}$ in these special events. Here, we used $\alpha = I / (\rho \cdot N_A / \mu) $ with an air density of $\rho = 1000$ kg/m on ground ($\rho = 100$ kg/m as proxy for high altitude), a molar mass of $\mu = 29 $ g/mol and the Avogadro constant $N_A= 6 \cdot 10^{23}$ / mol. So both ranges overlap, though the ionization rates in disks are more comparable to those on the ground level on Earth and below.

It might also be useful to have a look at the radiation at the Martian surface. The Radiation Assessment Detector onboard the Mars Science Laboratory (MSL) rover Curiosity on Mars measured about 200 $\rm \mu Gy / d$ \citep{Papaioannou2019}. This would be an equivalent dose of $\sim 2000 \mu Sv / d$,  assuming a weight factor of 10. So this radiation level is a few times higher than on the space station. 

For Venus \citet{Herbst2019} calculate a maximum atmospheric radiation dose for regular radiation conditions of 240 $\rm \mu Gy / d$ ($\sim 2400 \mu Sv / h$) in the upper atmosphere with non-extreme radiation events. 
Under regular conditions, the radiation in all these environments seems comparable to our applications in mind. 

Again, the similarities should not be pushed to the limit but at least this motivates an experiment based on cosmic radiation on Earth. Altitude dependent cosmic radiation should be a suitable first test case to evaluate the influence of radiation during planet formation.
We therefore built a balloon experiment that measures charging depending on height from ground to the stratosphere. The parameter of the surrounding gas shall be kept as constant as possible in order not to distort the results.

\begin{figure}
    \centering
	\includegraphics[width=\columnwidth]{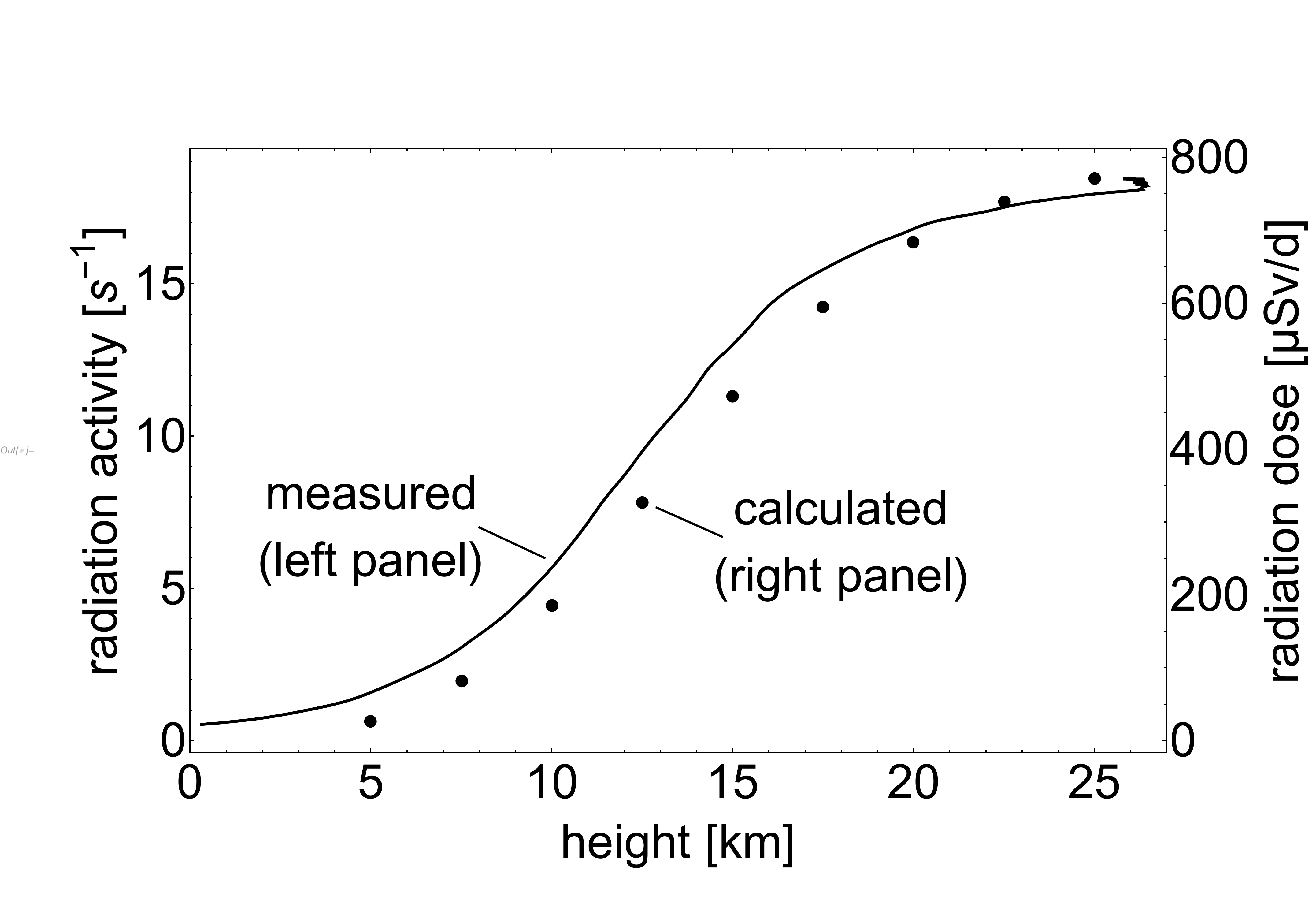}
	\caption{\label{fig.rad-height}Line: Measured radiation activity in dependence of the flight altitude, Dots: Calculated radiation dose based on the EPCARD code provided by Helmholtz Zentrum München}
\end{figure}

With this in mind, also radiation doses might be compared for different heights. Measurements on the space station show levels of 600 $\mu Sv/d$ (dose equivalent) \citep{Berger2017} while on ground one finds a typical dose equivalent of 6 $\mu Sv/d$. These measurements somewhat depend on the local environments and times but this is roughly a factor 100. In agreement to these general levels, we used the radiation calculator provided by the Helmholtz Zentrum München \citep{Mares2007} to give a height dependent radiation inside an aircraft as shown in fig. \ref{fig.rad-height}.

{The radiation levels off at about 700 $\mu Sv/d$ in the stratosphere, which would also be in agreement to the space station measurements at 400 km height not being higher.}
By using a balloon that is supposed to reach heights of 25-28 km, we therefore essentially sample the radiation levels from ground up to the space station. 

To monitor the radiation with height, we used a Geiger counter in the experiment. This only yields an activity value since the conversion factor for radiation above the ground is unknown. But in the context of this work we consider this sufficient as parameter to quantify the radiation level. The measurements of the experiment are also plotted in fig. \ref{fig.rad-height}. While the units are different, the profiles are very similar.

\section{Experiment}

The experiment flew under the acronym IROCS (Influence of Radiation on Charged Spheres) from Esrange (European Space and Sounding Rocket Range) in north Sweden together with three other experiments on 23.10.19 on the stratospheric balloon BEXUS 29. 

\subsection{Setup}

\begin{figure}
    \centering
	\includegraphics[width=\columnwidth]{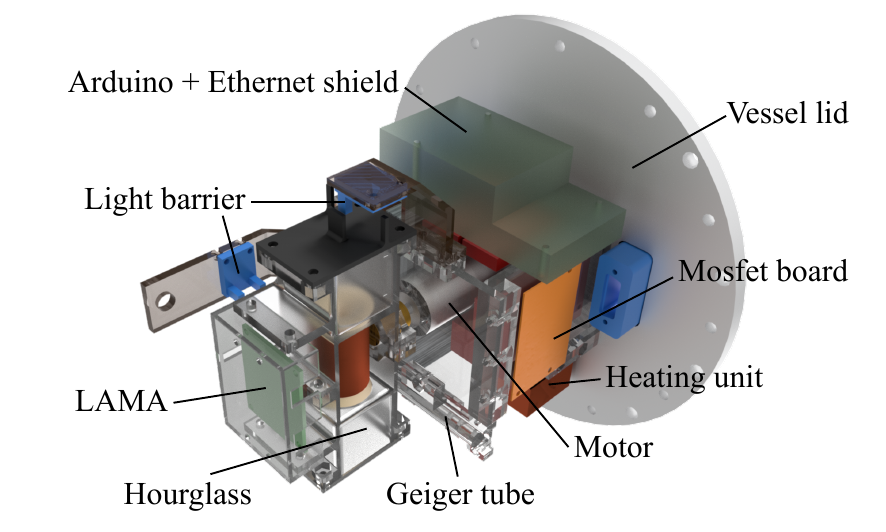}
	\caption{\label{fig.setup} Sketch of the experiment components mounted inside a pressure vessel. LAMA (LAdungsMess-Amplifier) refers to the charge amplifier. The hourglass as main part to charge and move grains is sketched in more detail in fig. \ref{fig.sketch}. The mosfet board controls the heating unit and power supply of the sensors.}
\end{figure}

The experiment is placed inside a 200 mm diameter, cylindrical, 3 mm thick aluminum pressure vessel. Bottom and top are 10 mm thick. The overall mass of the experiment is 8.5 kg. The vessel is air-tight to keep the pressure at 1 bar during the flight. Fig. \ref{fig.setup} shows a 3d sketch of the experiment itself without pressure vessel. The heating unit, consisting of a heating foil and a fan keeps the inside temperature at 20\textdegree C. To aid conditioning the pressure vessel is wrapped in a thermal insulation layer (not shown).

Inside the vessel all components  are mounted on an adapter made of PET, which is bolted to the inner vessel lid. The whole experiment is power supplied by batteries placed on top of the balloon's gondola.
An Arduino Mega microcontroller controls all actions and reads out the sensors. 
The temperature and pressure are measured inside and outside of the pressure vessel by two atmospheric sensors. The sensor measuring the outside pressure to calculate the balloon's altitude is placed outside on the lid between the vessel and one layer of insulation. To measure the radiation level a Geiger tube is used inside the vessel.

\begin{figure}
    \centering
	\includegraphics[width=\columnwidth]{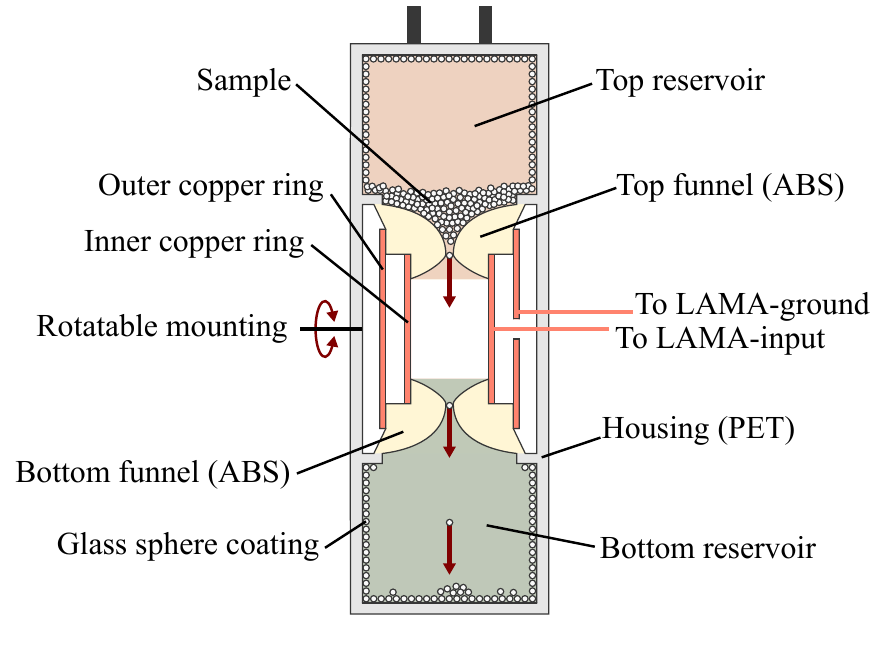}
	\caption{\label{fig.sketch} Sketch of the hourglass. There are two reservoirs of grains coated with the same glass spheres and combined by two copper rings. Grains falling from the top to the bottom reservoir pass through a copper ring. The induced charge is measured.}
\end{figure}

Fig. \ref{fig.sketch} shows a sketch of the experiment's main part, the hourglass-shaped charging unit. The total length of the hourglass is 143 mm. The hourglass can be rotated by a DC-motor and mainly consists of two grain reservoirs made of PET. The inner volume of each of the reservoirs is $35 \times 35 \times 35$ mm. The walls are coated by the glass spheres from the inside to avoid collisions between different materials, i.e. glass spheres and PET. The reservoirs are filled with about 2600 glass spheres with a diameter of $800 - 1000 \, \mu \text{m}$ and a grain mass of $1.02 \pm 0.16$ mg. Two light barriers are used to control the position and shaking of the hourglass.

The two reservoirs are connected by double-sided funnels and two concentric copper rings. To guarantee an undisturbed flow from the one reservoir to other these funnels are not coated by the glass spheres. The funnel orifice was chosen to be 2.5 mm in diameter to allow only single spheres to pass the funnel.

The inner copper ring (3 cm length, 2 cm outer diameter) is connected to a charge detector (LAMA) by the core of a coaxial cable going from the inner ring through a hole in the outer copper ring (see fig. \ref{fig.sketch}). To prevent the signal to be disturbed by  electromagnetic noise the outer copper ring (4 cm length, 3 cm outer diameter), the shielding of the coax cable and the housing of the LAMA is connected to the electrical ground of the power supply. The LAMA is placed rigidly on the hourglass to rotate together. This shall prevent the cables from twisting and thereby disturbing the signal. In general, the LAMA measures the electric potential difference between the ground (outer copper ring) and the inner copper ring. The spheres falling through the copper rings electrically induce a voltage depending on the charge, which is amplified by the LAMA.

This charge detector is a recent development \citep{genc2019,kaponig2020}. 
The output of the LAMA is fed to a 16 bit AD-converter and charges (voltages) are measured 860 times per second.

\subsection{Inflight operation}

By default, the experiment is fully operated by Arduino Mega 2560 software in an autonomous mode. It does not depend on a real time connection between the ground station and the experiment on the gondola and does not require manual intervention. 
An Ethernet shield is used for transmitting and saving the data on a 32 GB SD-card. Additionally, the Arduino serves as a server and accepts requests from the ground station. The ground station consists of a computer connected to the gondola via the E-Link system (telemetry). The sensor data and the state of the measuring cycle can be queried and changed if necessary. Furthermore, completed measurements can be downloaded from the SD card as a safety backup.

\begin{figure*}
    \centering
	\includegraphics[width=\textwidth]{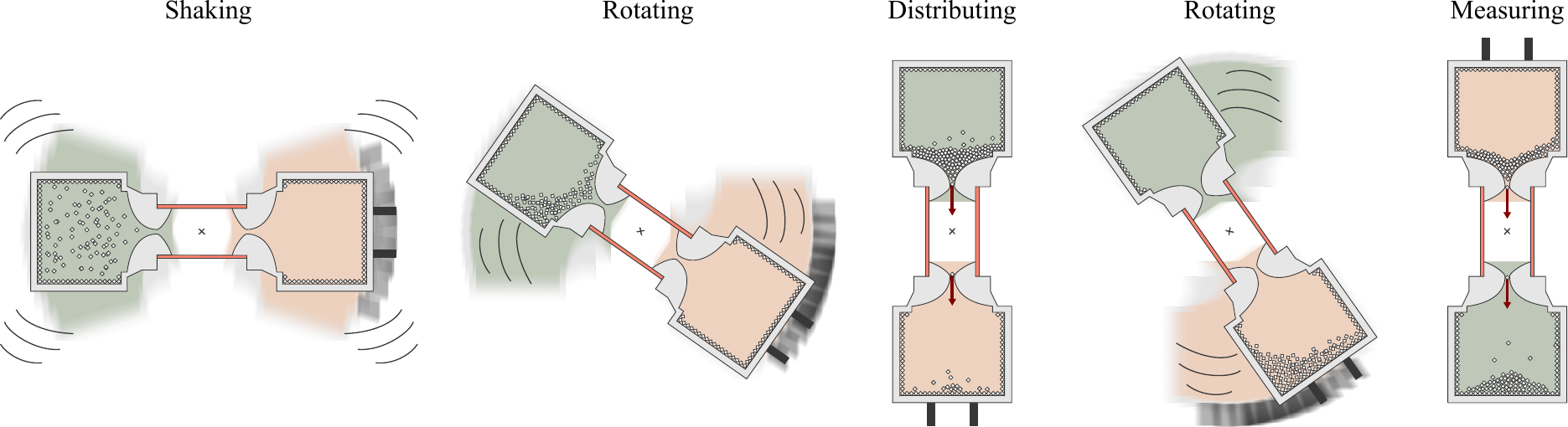}
	\caption{\label{fig.procedure} Measuring procedure: Shaking in horizontal position, rotating 90° and distributing grains to one side in vertical position, measuring charges after 180° rotation.}
\end{figure*}

Fig. \ref{fig.procedure}) sketches the whole measuring procedure. As the first step the hourglass is turned in the horizontal position and rotated about 15° up and down with a frequency of 3 Hz. This shaking procedure continues for 10 s which gives the radiation sensor time to perform a measurement. After that, the hourglass is rotated 90° in downright position and jolted for 7 s to distribute some spheres into the top reservoir. Meanwhile, the temperature sensor data are read out and the heating unit is adjusted. The hourglass is rotated 180° in the upright position, then the LAMA is switched on and the measurement is started. All spheres which got into the top reservoir during the distributing process are now stuck in the upper funnel. Providing some jolts to the hourglass single spheres are falling through the copper ring and the induced voltage is measured by the LAMA. In total, 10 jolts with an interval of 2 s are applied.

Then the cycle begins anew with shaking. 
In total, 174 cycles were performed during the flight and 35 closely before launch.

\subsection{Flight}

\begin{figure}
    \centering
	\includegraphics[width=\columnwidth]{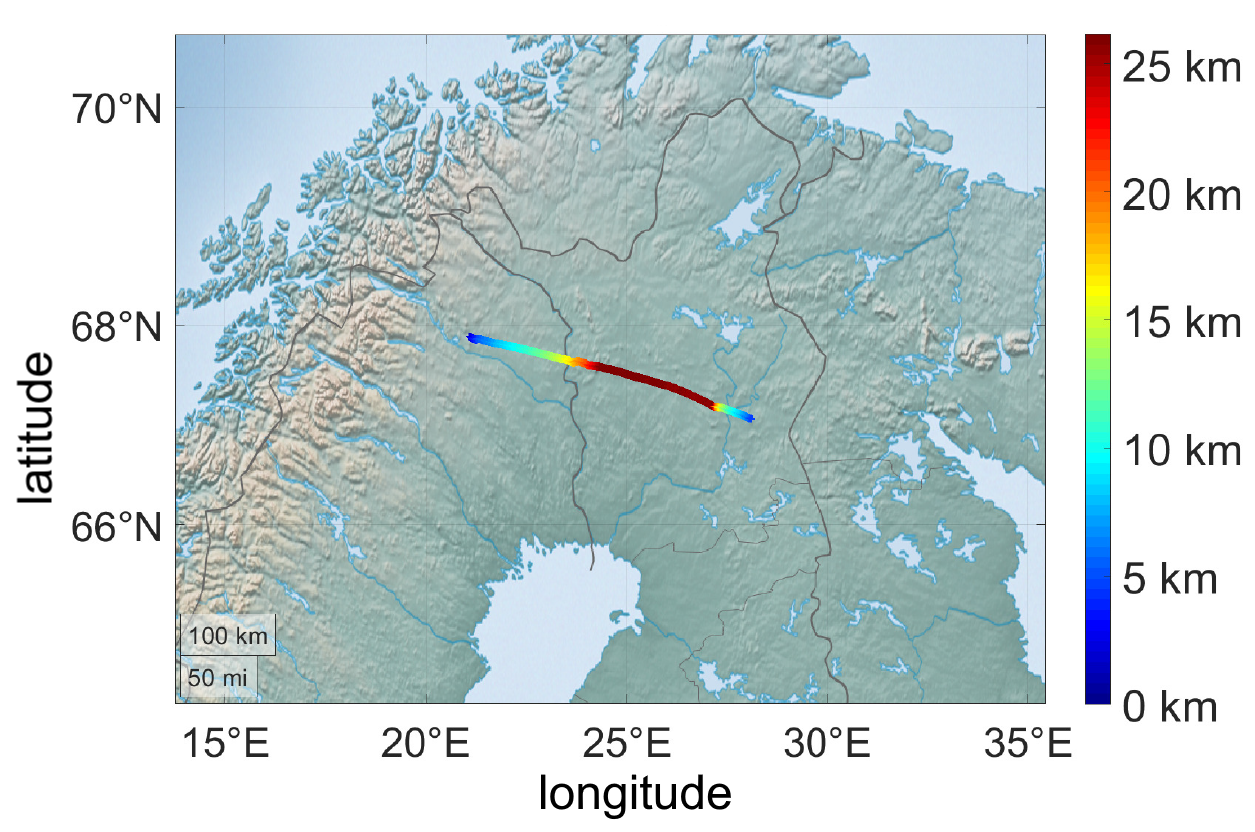}
	\caption{Flight trajectory of BEXUS 29. The altitude is color coded. The figure was generated with Matlab using open source maps.}
	\label{fig.karte}
\end{figure}

The balloon's trajectory is seen in fig. \ref{fig.karte}. The flight started at 06:33 local time an hour before dawn and had a duration of 3h and 26 min. After 1 h 25 min ascending time the balloon reached a maximum altitude of 26.14 km and floated close to that level for about one and a half hour. Coming close to the Russian border a cut-down of the balloon was initiated. Here the gondola was separated from the balloon and a parachute is opened to slowdown the descending of the gondola. After 29 minutes the gondola landed in north-east Finland about 320 km away from its launch position. 
In total, the flight lasted for 3 h and 26 min and provided a time of about 3 hours for experiments, since the measurement cycle was stopped shortly before the cut-down. Atmosphere data from inside and outside of the vessel were measured until the power supply was removed from the experiment during the gondola's recovery.

Charging might depend on the temperature and the pressure. We, therefore, tried to keep both quantities as constant as possible. 
 
\begin{figure}
    \centering
	\includegraphics[width=\columnwidth]{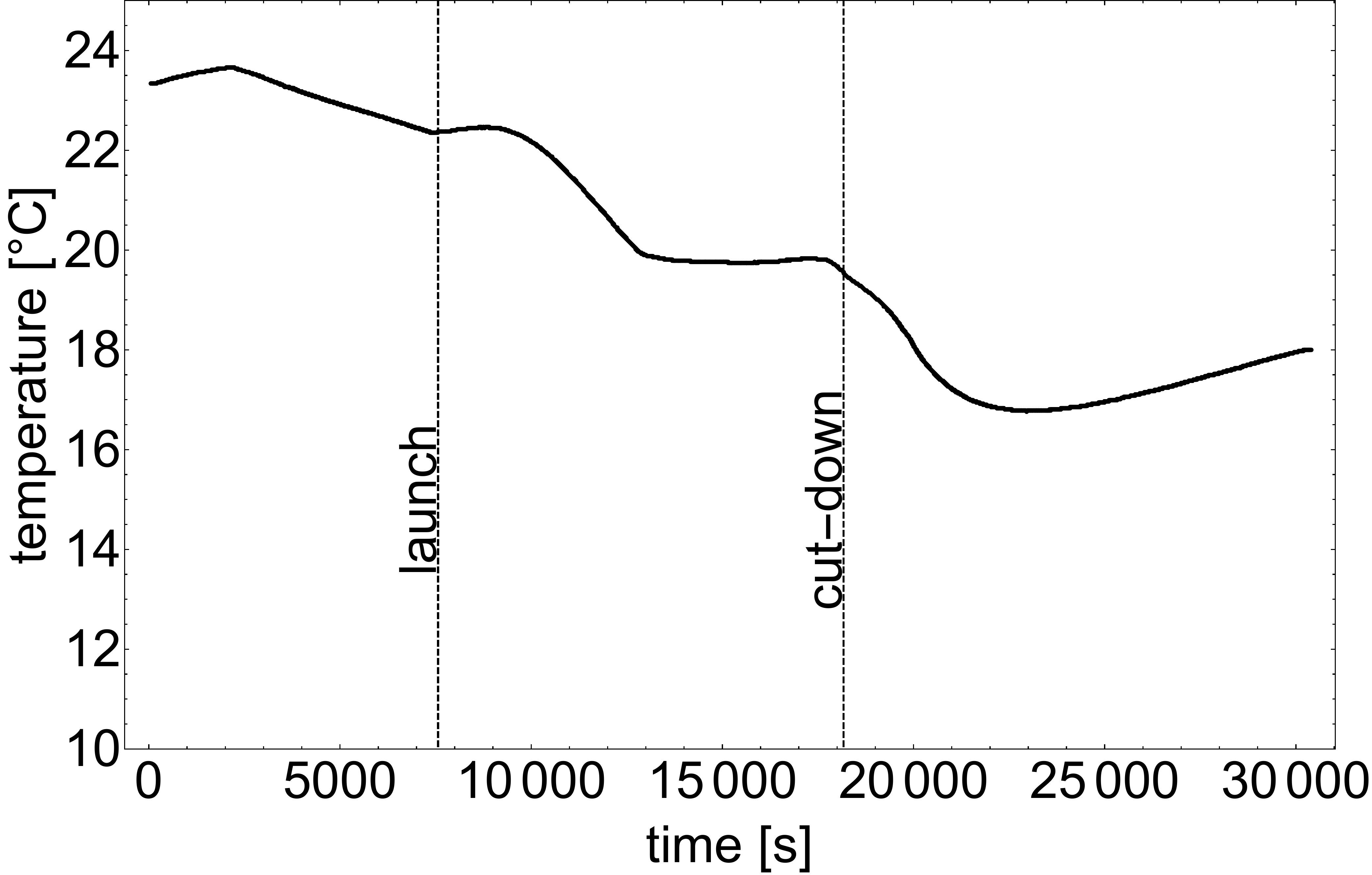}
	\caption{Temperature inside the pressure vessel. Launch time and end of the floating phase are marked by vertical lines.}
	\label{fig.temp}
\end{figure}

The temperature profile is shown in figure \ref{fig.temp}. Depending slightly on the temperature the relative air humidity inside the vessel varies in a range from $34-36 \%$.

The measured pressure inside the airtight vessel 

is nearly constant at 950 mbar. Small fluctuations are caused by minor temperature differences in the progress of the flight.

\section{Results}
\subsection{Radiation levels}

Before each charge measurement, the Geiger tube integrated the events caused by radiation for 10 s.
Fig. \ref{fig.rad-height} shows the dependency of height (calculated from the outside pressure) and the measured mean radiation levels. Keeping in mind that the radiation is measured as activity, the profile is similar to the calculated radiation dose in fig. \ref{fig.rad-height}. Before launch and during floating phase the radiation activity is at a constant level. At the ground the rate is 0.33 $s^{-1}$ and during floating phase 18.1 $s^{-1}$, therefore we measured an increase of a factor of 54.

\subsection{Collisional charging}

\begin{figure}%
    \centering
{{\includegraphics[width=\columnwidth]{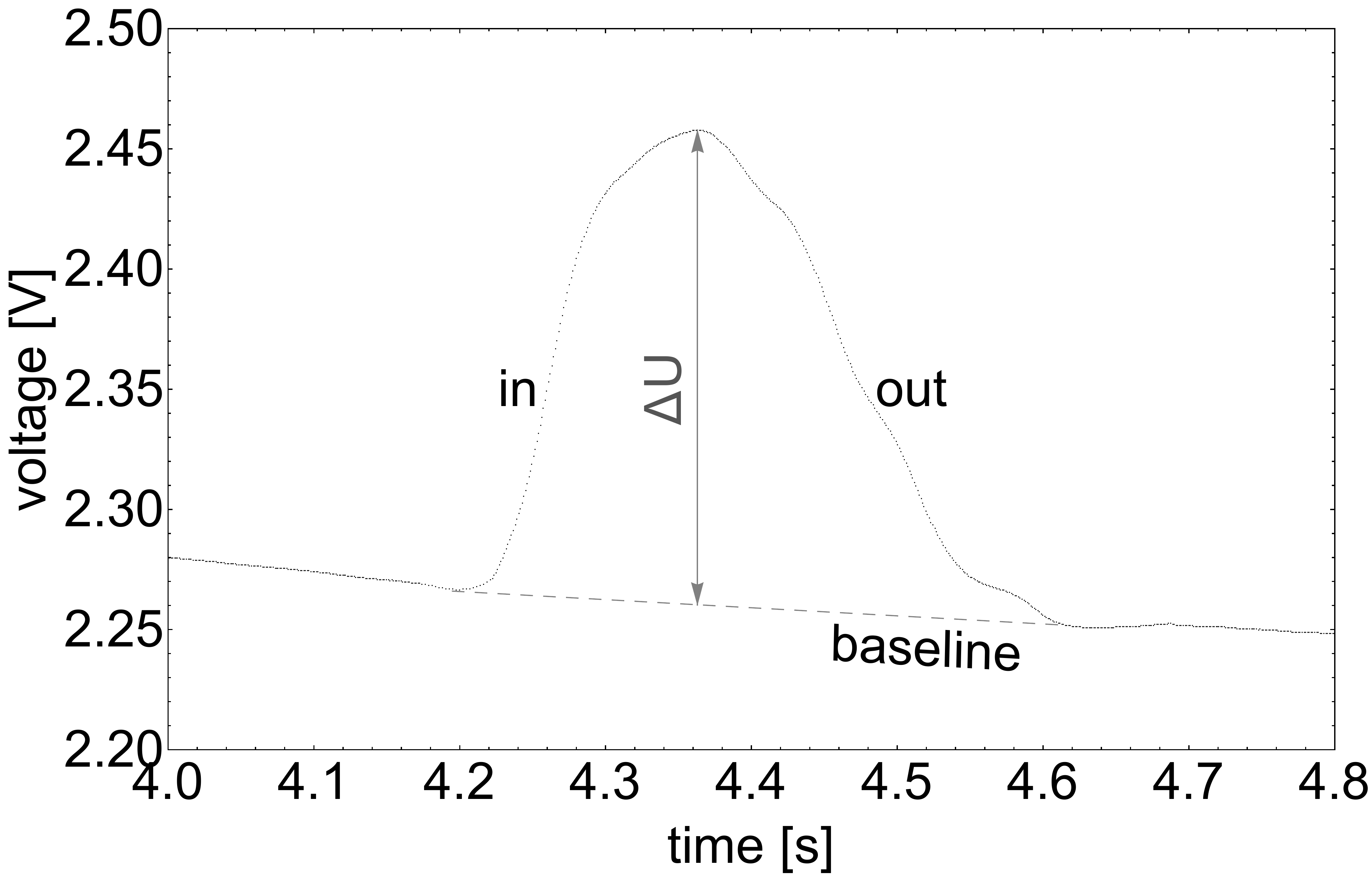} }}%
    \qquad
{{\includegraphics[width=\columnwidth]{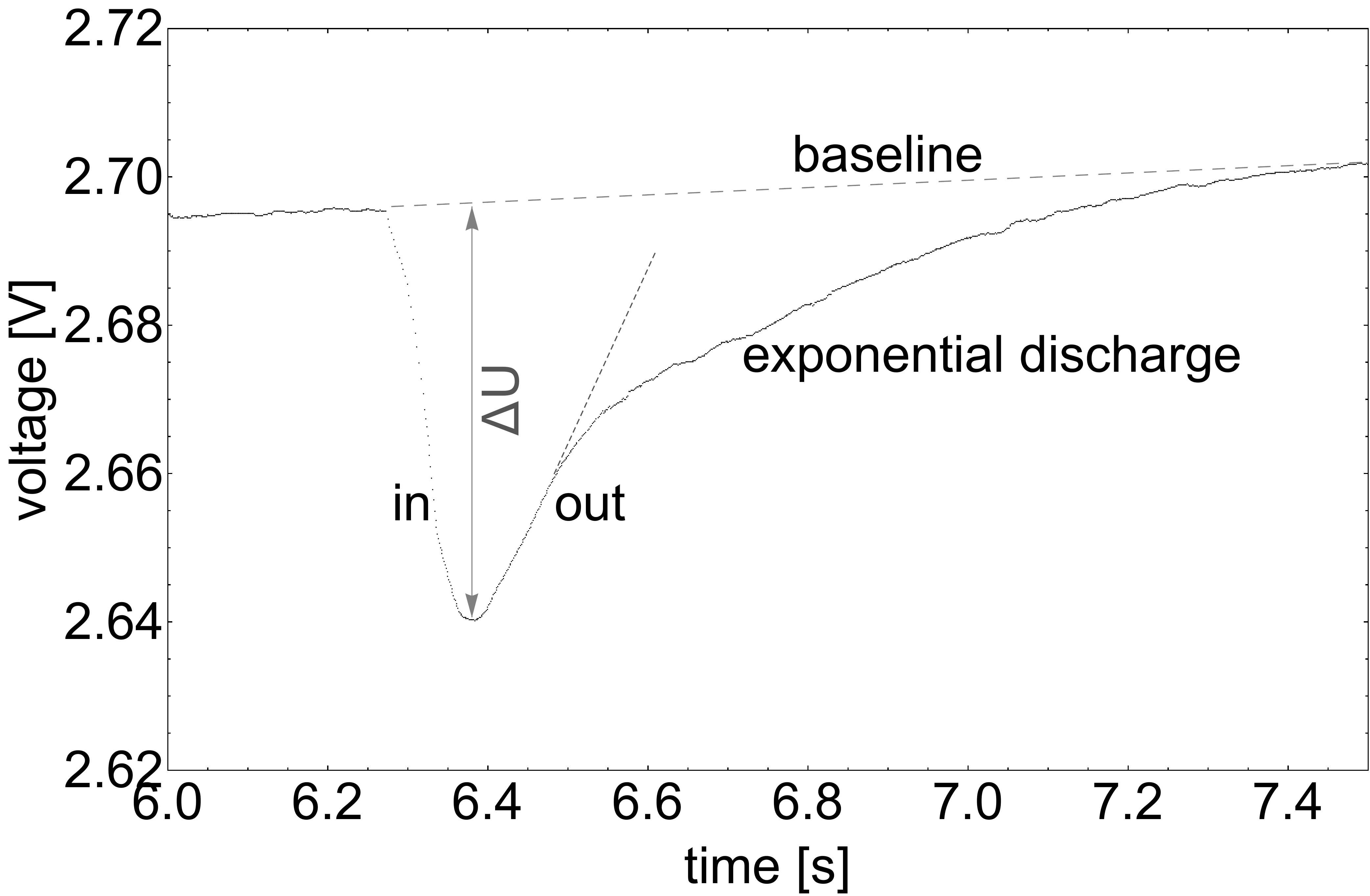} }}%
    \caption{\label{fig.charges}Signals of single charged spheres falling through the copper rings inducing a voltage measured by the LAMA; top: positive particle; bottom: negative particle, the long tail is attributed to charge transferred to the bottom funnel, slowly discharging due to connection to the ground. Typical signal lengths are 0.3 - 0.4 s because the spheres are not falling straight through the copper (0.1 s free fall time).}%
\end{figure}

During every measurement cycle, a voltage curve is measured by the LAMA. From these data signals are extracted originating from a single charged sphere passing through the copper rings. Examples for a negative particle as well as a positive charged sphere are shown in figure \ref{fig.charges}.

When a jolt of the motor occurs a sphere is falling through the funnel and enters the copper ring. If the sphere is positively charged the induced voltage is increasing. At the maximum, the sphere reaches the geometric centre of the copper ring. The sensor drifts slightly. This is taken into account to calculate the baseline of an event and the maximum voltage induced. As the sphere leaves the copper ring, the voltage decreases again and the voltage reaches its original value as shown for a positive grain in fig. \ref{fig.charges} (top). We attribute the slight asymmetries of the signal and the signal length of 0.3 - 0.4 s to details of the trajectories. Grains enter through the funnel, where collisions decelerate the grains. These contacts of only milliseconds do not change the charge state of the grain significantly compared to the high number of collisions during charging but explain the timescale of the signal which is somewhat slower than free fall. 

The charge can be calculated by measuring $\Delta U$. Due to the finite length of the electrodes and open top and bottom, the absolute charge is  underestimated and the signal does not reach a plateau. This occurs systematically though for all charges and does not hinder a further comparison of charges relative to each other. 

\begin{figure}
    \centering
	\includegraphics[width=\columnwidth]{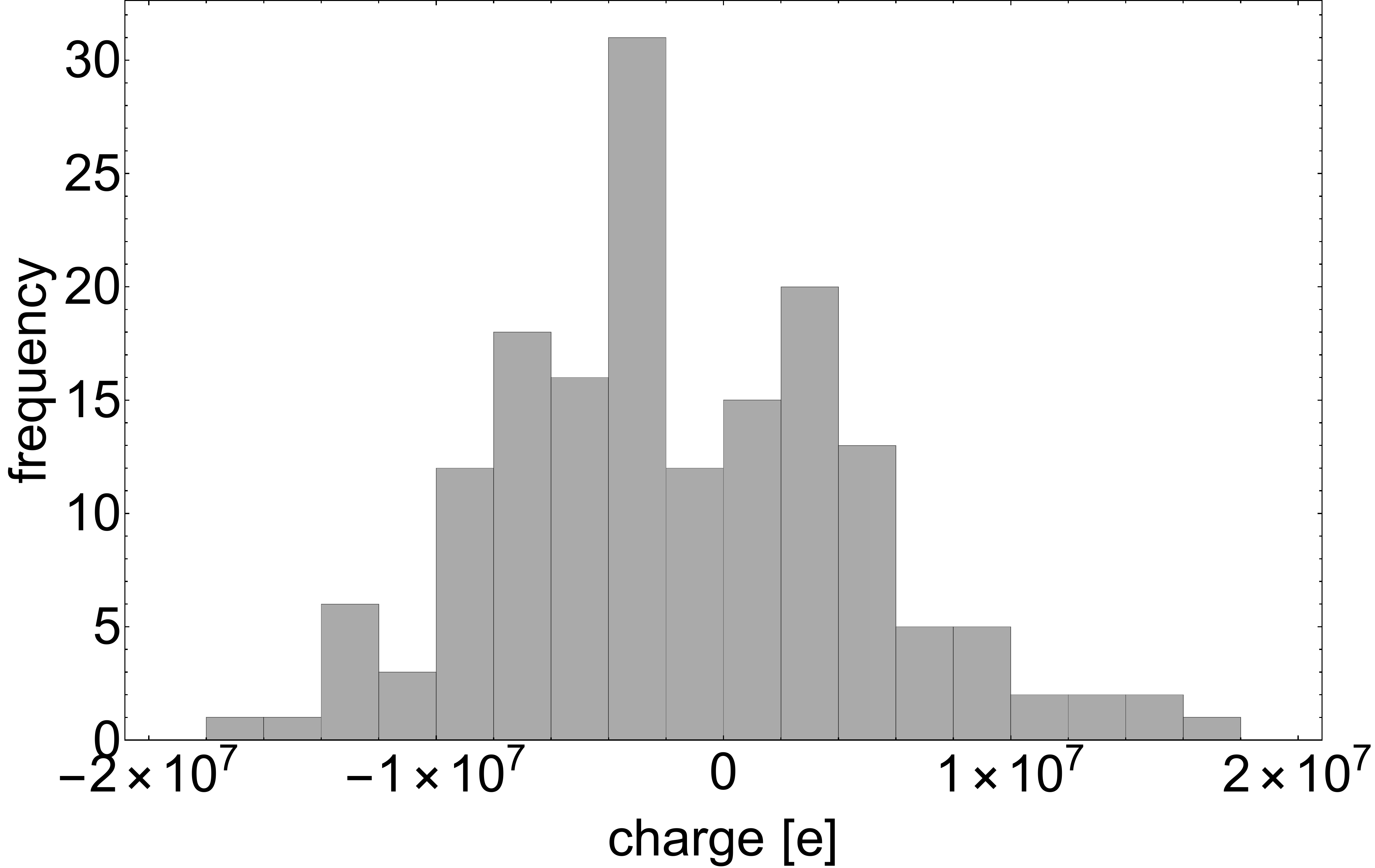}
	\caption{\label{fig.all-histo} All 170 measured charges during flight and pre-flight experiments. Due to particle-particle charging of identical grains there are positive as well as negative charges.}
\end{figure}

Fig. \ref{fig.charges} (down) shows the signal of a negatively charged sphere with a slightly different structure. While the electrode recovers as the grain approaches the bottom, there is a kink in the signal. We attribute this to the sphere touching the walls of the funnel for a longer timescale and transferring some charge. 
The signal then merges into an exponential recovery. This can be explained by a slow discharge of the funnel to the copper ring. The timescale of discharge (RC) with a capacitance of $\approx$ pF and a resistance of $\approx$ T$\Omega$ of the insulator is a reasonable match to the observed exponential. 
These signals are not the rule as a significant amount of charge can only be transferred in the case a sphere gets to rest on the ABS.
Grains which get in direct contact with the funnel during a collision, do not imply that the spheres discharge in general.
In addition, while the ABS has a significant conductivity in the context of this experiment, the glass of the sample grains does not. All container walls which get in contact with the sample grains are covered with the same glass particles, which serves as an insulating layer and collisional charging occurs only due to collisions with identical grains. Glass particles do not discharge on the timescales of minutes important in this experiment \citep{Jungmann2018, Steinpilz2020a}. In contrast, they even keep complex charge patterns on their surface \citep{Steinpilz2020b}. Also, a charging between grains and ABS would result in a bias towards negative or positive charges on the grains due to material differences along the triboelectric material order. As seen in fig. \ref{fig.all-histo}, there is no significant bias. Any short contact between grains and ABS, e.g. during passage through the Faraday tube, only changes the total charge slightly.
In any case, $\Delta U$ represents the initial charge of the sphere up to a correction factor which is the same for all grains. Due to the open electrodes, measurements only give an estimated charge which is systematically too low. Again, as we only compare the charges relatively to each other, this does not affect the overall conclusion.

The net charge of a sphere $Q$ can be estimated as
\begin{equation}
    Q=\dfrac{C}{A} \cdot \kappa \cdot \Delta U.
\end{equation}

Here $C=63,3 \,\text{pF}$ is the capacity of the LAMA, copper ring and wires, $A=11$ is the factor by which the voltage is amplified by the LAMA, and $\kappa=2.434$ is the factor of a voltage divider. \\

\begin{figure}
    \centering
	\includegraphics[width=\columnwidth]{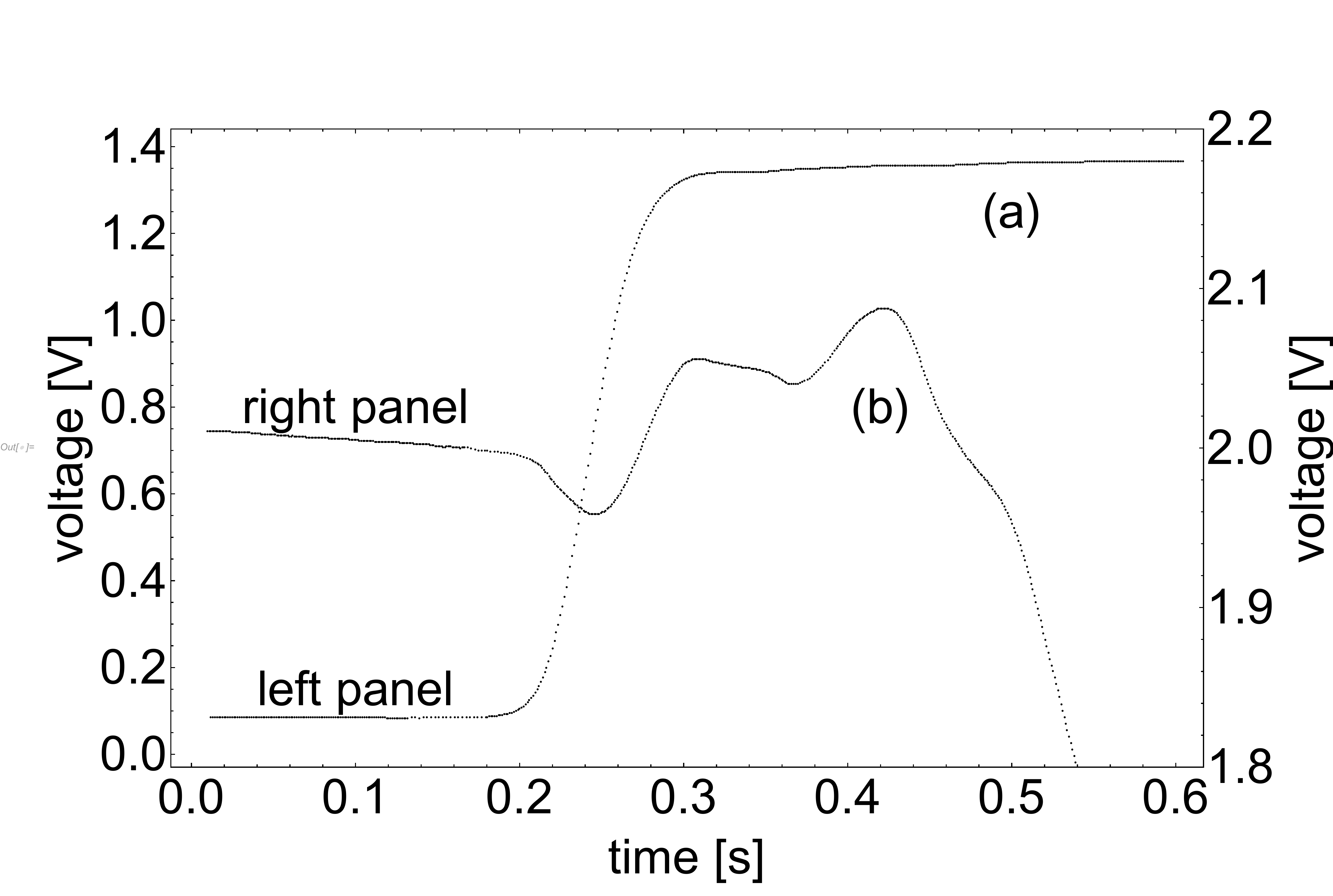}
	\caption{\label{fig.not-useable}Signals discarded for charge evaluation. (a) large voltage jumps induced by a disturbance, (b) signal of multiple spheres oppositely charged}
\end{figure}

We note that only some signals occurring after the jolts can be used to evaluate a charge of a sphere. If there are several spheres following each other too closely, the signals overlap and an unequivocal correlation of spheres and the dips/peaks of the voltage signal is no longer possible. An example is shown in fig. \ref{fig.not-useable} (b). As the electronic is very sensitive, also abrupt jumps of the voltage can occur as shown in fig. \ref{fig.not-useable} (a). Such events are ignored as well. 

Fig. \ref{fig.all-histo} shows a histogram of all measured charges.
The order of magnitude of maximum charge of $10^7$e is consistent with earlier measurements of charge densities (charge per surface area) on grains made of the same material in other ground based or drop tower experiments \citep{Jungmann2018, Steinpilz2020a}. This indicates that the charging process though resulting from a different excitation of the grains yields the same equilibrium values. The presence of positive as well as negative charges emphasizes an underlying stochastic origin  \citep{haeberle2018}.

\begin{figure}
    \centering
	\includegraphics[width=\columnwidth]{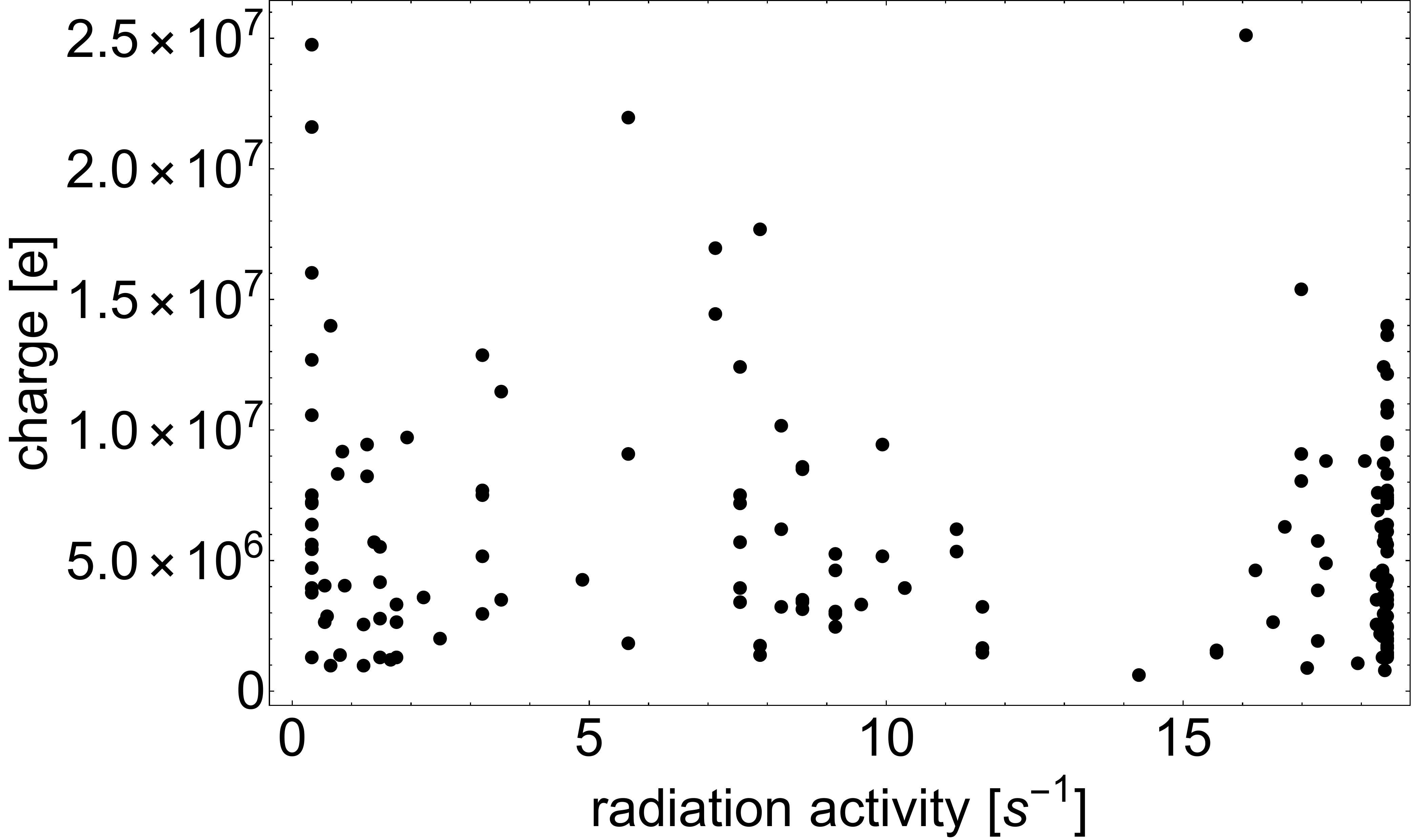}
	\caption{\label{fig.all-charge}Absolute values of measured charges over radiation level. Due to the required quality of the signals, data are not distributed evenly during ascent which results in two regions less populated with data points.}
\end{figure}

Fig. \ref{fig.all-charge} shows the absolute charges on each grain at the respective radiation level. The width mirrors the natural charge distribution. The data are showing three distinct groups. Values at low radiation are mostly ground values and early ascend, low altitude values. The second group gives charges at the fast ascending phase. The third group has many data points due to the long floating phase at maximum height.

\begin{figure}
    \centering
	\includegraphics[width=\columnwidth]{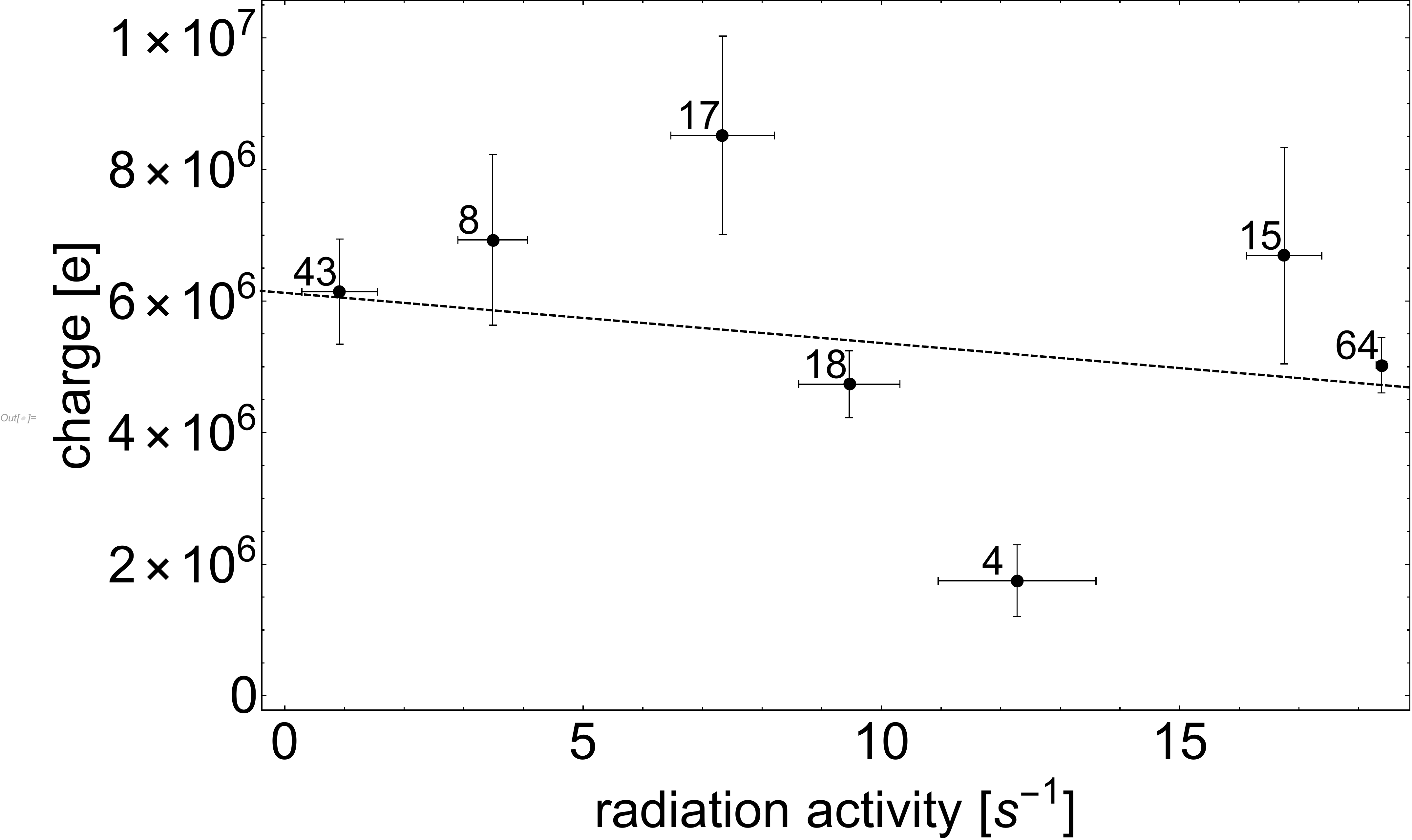}
	\caption{\label{fig.bin3}Charges averaged for equidistant radiation rates of 3 $\rm s^{-1}$ each. Error bars in charge are average errors or the mean. Error bars in radiation are standard deviations. The number of measurements of each data point is marked. The line marks a linear trend consistent with the error bars, not considering one outlier which we attribute to the low number of included data.}
\end{figure}

The radiation dependent charge values are shown in fig. \ref{fig.bin3}. Here, data for 3 $\rm s^{-1}$ intervals are averaged.

There is one outlier which we attribute to the small number of 4 measurements entering here. Otherwise, within the error bars, there is a slight decrease between the first and last data point. We included a linear trend consistent with the error bars. In any case, the slope is very shallow and the maximum possible decrease is only about 30 \% within the parameter range.

\section{Discussion}

Grain charges generated by collisions are up to about $10^7$e on the ground. At a radiation dose 54 times larger the charge decreases measurably but only 30\%. 

Discharge by free ions cannot explain this reduction.
As estimated above, radiation only produces about 1000 free ions per $cm^{3}$ per second. As worst case, this is used completely to discharge grains. On the charging timescale of minutes or $10^2\,$s this is a total of $10^5$ charges in the experiment volume. Distributed equally to $10^3$ grains this is $10^2$ charges per grain, which is negligible in view of up to $10^7$ e produced by collisional charging on a grain and continuous recharging.

The decrease in charging is only marginal. Therefore, also small changes in common environmental parameters might be responsible. It is known that charging depends on humidity \citep{Schella2017}. Especially, high levels of humidity might suppress charging. While  humidity in the experiment only varies slightly over time, we cannot rule out an influence on the small scale measured. Especially, as the temperature decreases (though only slightly), the humidity increases in agreement to lower charge levels at later times.

\section{Conclusion}

Cosmic radiation ionizes its environment and might be important for all processes that involve charges in one way or the other. In a balloon-borne experiment, we collisionally charged mm-sized spherical glass particles and measured their charge. The only parameter varying strongly during the experiment was cosmic radiation, which increased with altitude by a factor of 54 in rate measured by a Geiger sensor. 

Only a very small decrease in charging from ground to the stratosphere was observed.

On the downside, with this small difference in charging measured, we cannot decide, if radiation is responsible at all as we cannot exclude the influence of slight effects by temperature and humidity. 
On the bright side, 
the differences are so small in any case, that cosmic radiation might fairly be ignored during collisional charging on timescales of minutes.

That does not mean that environments might not change on long timescales by cosmic radiation. However, it implies that -- on the level of atmospheric dose rates on Earth, Mars and in protoplanetary disks -- radiation plays no immediate role during collisional charging, does not lead to discharge at least on timescales of minutes and therefore is not important in any short time, charge related events like thunderstorms on Earth or charging in dust devils on Mars.

\section*{acknowledgements}
We thank the German Aerospace Center (DLR), the Swedish National Space Agency (SNSA) and the European Space Agency (ESA) for the granting access to a stratospheric balloon within the REXUS/BEXUS-program. F. Jungmann acknowledges support by DLR with funds provided by the BMWi under grant 50WM1762, the project that did set the stage for the balloon experiment. N. Schneider is supported by funds provided by the German research foundation under grant WU 321/16-1. We thank the Förderverein Universität Duisburg-Essen for travel funds. We also appreciate the continuous support of all kinds by ZARM-FAB during the project, i.e. Simon Mawn, Dieter Bischoff and Torsten Lutz. We also thank the Mobile Rocket Base (Moraba), the Swedish Space Corporation (SSC) and especially Stefan Krämer for the successful balloon campaign.

\nolinenumbers

\bibliography{bib}

\end{document}